\documentclass[review,10pt]{elsarticle}
\usepackage{amsmath,amsthm,amsfonts,amssymb,mathtools,color,xcolor}
\usepackage[hidelinks]{hyperref}
\hypersetup{colorlinks, linkcolor=blue, citecolor=blue}
\journal{IJB}
\begin{document}
\begin{frontmatter}

\title{Approximate Analytical Solution of a Cancer Immunotherapy Model by the Application of Differential Transform and Adomian Decomposition Methods}

\author{Alireza Momenzadeh$^1$}
\author{Sima Sarv Ahrabi$^{2,}$ \corref{cor1}}
\ead{sima.sarvahrabi@sbai.uniroma1.it}
\cortext[cor1]{Corresponding author}

\begin{abstract}
Immunotherapy plays a major role in tumour treatment, in comparison with other methods of dealing with cancer. The Kirschner-Panetta (KP) model of cancer immunotherapy describes the interaction between tumour cells, effector cells and interleukin--2 which are clinically utilized as medical treatment. The model selects a rich concept of immune-tumour dynamics. In this paper, approximate analytical solutions to KP model are represented by using the differential transform and Adomian decomposition. The complicated nonlinearity of the KP system causes the application of these two methods to require more involved calculations. The approximate analytical solutions to the model are compared with the results obtained by numerical fourth order Runge-Kutta method.
\end{abstract}

\end{frontmatter}

\section{Introduction}\label{sec.intro}

In past years, cancer immunotherapy has considered as a method of enhancing the features of cancer treatment. Immunotherapy is mentioned as the utilization of synthetic and natural substances in order to stimulate the immune response. This could be achieved in clinical treatments by using immunostimulators such as interleukins, cell growth factors, T cells and so forth. T cell is a type of lymphocyte, which performs a major role in cell-mediated immunity. The immune response to cancer cells is normally cell-mediated, therefore, T cells and natural killer cells are the centre of attention. Medical research indicates that a clinical practice, which consists in using interleukins leads to the stimulation of immune responses. Interleukin-2 (IL-2) is the principal cytokine, which regulates white blood cells and is mainly produced by CD4\textsuperscript{+} T cells.

The dynamics of tumour-immune interaction have been studied over the past years and several theoretical models have been developed by researchers to indicate and analyse the influence of immune system and tumour on each other. In \cite{kuznet94}, the authors have presented a mathematical model involving ordinary differential equations describing the T lymphocyte response to the growth of an immunogenic tumour. Adam \cite{adam96} has developed and analysed a system consisting of two ordinary differential equations, which represents the effect of vascularization within a tumour. In \cite{deBoer85} the authors have introduced some detailed models including 8--11 differential equations and 3--5 algebraic equations to illustrate the T lymphocyte interactions, which generates anti--tumour immune response. Kirschner and Panetta \cite{kirsch98} have investigated the cancer dynamics and presented a model, which richly describes the interaction between the effector cells, the tumour cells and the concentration of IL-2 and addresses long-term and tumour recurrence and short-term tumour oscillations. Banerjee and Sarkar \cite{baner08} have enhanced a system of delay ordinary differential equations to describe the reciprocal interaction between tumour, T-lymphocytes and T-helper cells. More mathematical models, incorporating delay and stochastic models could be observed in \cite{eftim11}.

The Kirschner--Panetta (KP) Model, which was first introduced in \cite{kirsch98}, has selected rich immune-tumour dynamics, but nevertheless remains as straightforward as possible whilst incorporating crucial factors of cancer immunotherapy. The rich and involved nonlinearity of the system is generated by three terms of Michaelis–-Menten type, hence analytical approaches in solving the nonlinear system require more complicated processes. Analytical expression of approximate solution to the system is represented by using the differential transform method (DTM) and also Adomian decomposition method (ADM). The next sections are allocated to briefly describe the mentioned methods and compare obtained results.

\section{Kirschner-Panetta model}\label{sec.KP}

the KP model \cite{kirsch98} indicates the dynamics of immune-cancer by defining three populations, namely $E\left(t\right)$, the effector cells such as cytotoxic T-cells; $T\left(t\right)$, the tumour cells; and $I\left(t\right)$, the concentration of IL-2:
\begin{align}
  \label{eq.kp1}
  \frac{\mathrm{d}E}{\mathrm{d}t}&=cT-{{\mu }_{2}}E+\frac{{{p}_{1}}E{{I}}}{{{g}_{1}}+{{I}}}+{{s}_{1}}\,,\\
  \label{eq.kp2}
  \frac{\mathrm{d}T}{\mathrm{d}t}&={{r}_{2}}T\left( 1-bT \right)-\frac{aET}{{{g}_{2}}+T}\,,\\
  \label{eq.kp3}
  \frac{\mathrm{d}{{I}}}{\mathrm{d}t}&=\frac{{{p}_{2}}ET}{{{g}_{3}}+T}-{{\mu }_{3}}{{I}}+{{s}_{2}}\,,
\end{align}
with the initial conditions:
\begin{equation}\label{eq.ini}
  E\left(0\right)=E_0 \,, \quad T\left(0\right)=T_0 \,, \quad I\left(0\right)=I_0 \,.
\end{equation}
The Eq. \eqref{eq.kp1} represents the rate of change in effector cells. The first and third term on the right-hand side of \eqref{eq.kp1}, show the stimulation of effector cells. The parameter $c$ indicates the immunogenicity of the tumour, i.e. the ability of the tumour to provoke an immune response. The third term, which is the model of Michaelis–Menten kinetics shows the saturated effects of the immune response. The parameter $s_1$ represents an external source of effector cells as medical treatments. The parameter $\mu_2$ indicates the decay rate of the effectors. The natural lifespan of the effector cells is in fact $\frac{1}{\mu_2}$ days. The rate of change of the tumour is described in Eq. \eqref{eq.kp2}. The tumour growth is described in terms of a limiting-growth model, i.e.\ logistic growth with the carrying capacity equal to $1/b$. The parameter $a$ represents the ability of the immune system to resist the tumour, i.e. the rate of interaction between tumour and effector cells. The third equation in the model, Eq. \eqref{eq.kp3}, indicates the rate of change of IL-2 concentration. The first term on the right-hand side of Eq. \eqref{eq.kp3} has Michaelis-Menten kinetics and illustrates that effector cells are stimulated by the interaction with tumour cells and therefore this interaction is a source of IL-2. The decay rate of interleukin-2 is expressed by the parameter $\mu_3$ and finally $s_2$ is a source of IL-2 for medical treatments. The values of the parameters in Eqs. \eqref{eq.kp1}-\eqref{eq.kp3} are given in Table~1 (see \cite{kirsch98}). The units of $g_1$, $g_2$, $g_3$ and $b$ are volume and the units of the other parameters are $day^{-1}$.
\begin{table}[h!]
\centering
\begin{tabular}{ |p{3.4cm}|p{3.4cm}|p{3.4cm}|  }
 \hline
 \multicolumn{3}{|c|}{\textbf{Values of parameters}} \\
 \hline
 Parameters in Eq. \eqref{eq.kp1}         &    Parameters in Eq. \eqref{eq.kp2}      &     Parameters in Eq. \eqref{eq.kp3}\\
  \hline
$0\le c\le 0.05$\hphantom{=} & ${r}_{2}=0.18$\hphantom{=} & ${p}_{2}=5$\hphantom{=} \\
${\mu }_{2}=0.03$\hphantom{=} & $b=1\times {10}^{-9}$\hphantom{=} & ${g}_{3}=1\times {10}^{3}$\hphantom{=} \\
${p}_{1}=0.1245$\hphantom{=} & $a=1$\hphantom{=} & ${\mu }_{3}=10$\hphantom{=} \\
${g}_{1}=2\times {10}^{7}$\hphantom{=} & ${g}_{2}=1\times {10}^{5}$\hphantom{=} &\\ 
 \hline
\end{tabular}
\end{table}

The model introduced by Kirschner and Panetta \cite{kirsch98} is a stiff system of ordinary differential equations, since a very small disturbance in time results in very large changes in some of the variables. Thus, without an appropriate scaling, the prevalent numerical methods of solving differential equations might fail. In order to normalize the model, the following scaling could be utilized: $x=E/{E}_{0}$, $y=T/{T}_{0}$ and $z=I/{I}_{0}$. The non-dimensionalized coefficients are given in Table~2.
\begin{table}[ht!]
\centering
\begin{tabular}{ |p{3.4cm}|p{3.4cm}|p{3.4cm}|  }
 \hline
 \multicolumn{3}{|c|}{\textbf{Scaled Coefficients}} \\
 \hline
 Coefficients in Eq.\eqref{eq.kp1}         &    Coefficients in Eq.\eqref{eq.kp2}      &     Coefficients in Eq.\eqref{eq.kp3}\\
  \hline
$\bar{c}=\displaystyle \frac{ c{{T}_{0}}}{{{E}_{0}}}$\hphantom{=} & ${{\bar{r}}_{2}}=\displaystyle {{r}_{2}}$\hphantom{=} & ${{\bar{p}}_{2}}=\displaystyle \frac{{{p}_{2}}{{E}_{0}}}{{{I}_{0}}}$\hphantom{=} \\
$\bar{\mu }_{2}={\mu }_{2}$\hphantom{=} & $\bar{b}=b T_0$\hphantom{=} & $\bar{g}_{3}=\displaystyle \frac{g_3}{T_0}$\hphantom{=} \\
$\bar{p}_{1}={p}_{1}$\hphantom{=} & $\bar{a}=\displaystyle \frac{ a{{E}_{0}}}{{{T}_{0}}}$\hphantom{=} & $\bar{\mu }_{3}={\mu }_{3}$\hphantom{=} \\
$\bar{g}_{1}=\displaystyle \frac{g_1}{I_0}$\hphantom{=} & $\bar{g}_{2}=\displaystyle \frac{g_2}{T_0}$\hphantom{=} & $\bar{s }_{2}=\displaystyle \frac{s_2}{I_0}$\hphantom{=} \\
$\bar{s }_{1}=\displaystyle \frac{s_1}{E_0}$\hphantom{=} &  &\\
 \hline
\end{tabular}
\end{table}

By eliminating the overbar notation, the scaled model is obtained as follows:
\begin{equation}\label{eq.kpScaled}
  \left\{ \begin{array}
  {r@{\quad=\quad}l}
   \displaystyle \frac{\mathrm{d}x}{\mathrm{d}t} & cy-{{\mu }_{2}}x+\displaystyle \frac{{{p}_{1}}xz}{{{g}_{1}}+z}+{{s}_{1}} \,, \\
   \displaystyle \frac{\mathrm{d}y}{\mathrm{d}t} & {{r}_{2}}y\left( 1-by \right)-\displaystyle \frac{axy}{{{g}_{2}}+y} \,, \\
   \displaystyle \frac{\mathrm{d}z}{\mathrm{d}t} & \displaystyle \frac{{{p}_{2}}xy}{{{g}_{3}}+y}-{{\mu }_{3}}z+{{s}_{2}} \,,
\end{array} \right.
\end{equation}
with the initial condition
\begin{equation}\label{eq.iniScaled}
  x\left(0\right)=x_0 \,, \quad y\left(0\right)=y_0 \,, \quad z\left(0\right)=z_0 \,.
\end{equation}
Since the carrying capacity of the tumor is equal to $1/b=10^{9}$, one reasonable scaling is to define ${E}_{0}={T}_{0}={I}_{0}=10^{4}$. In this case, the values for scaled parameters is represented in Table~3.
\begin{table}[h!]
\centering
\begin{tabular}{ |p{3.4cm}|p{3.4cm}|p{3.4cm}|  }
 \hline
 \multicolumn{3}{|c|}{\textbf{Values of scaled parameters}} \\
 \hline
 Parameters in Eq. \eqref{eq.kp1}         &    Parameters in Eq. \eqref{eq.kp2}      &     Parameters in Eq. \eqref{eq.kp3}\\
  \hline
$0\le c\le 0.05$\hphantom{=} & ${r}_{2}=0.18$\hphantom{=} & ${p}_{2}=5$\hphantom{=} \\
${\mu }_{2}=0.03$\hphantom{=} & $b=1\times {10}^{-5}$\hphantom{=} & ${g}_{3}=0.1$\hphantom{=} \\
${p}_{1}=0.1245$\hphantom{=} & $a=1$\hphantom{=} & ${\mu }_{3}=10$\hphantom{=} \\
${g}_{1}=2\times {10}^{3}$\hphantom{=} & ${g}_{2}=10$\hphantom{=} &\\
 \hline
\end{tabular}
\end{table}

\section{Differential transform and Adomian decomposition methods}\label{sec.DTMADM}

As it is obvious from Eq. \eqref{eq.kpScaled}, the complicated nonlinearity of the KP model causes the utilization of methods such as DTM and ADM to require more involved processes. In this section, for the sake of brevity, the basic concepts of DTM and ADM are discussed, then these two methods are utilized to present approximated analytical solution to the KP model.

\subsection{Differential transform method}\label{sec.DTMADM.DTM}

The differential transform technique was first introduced in \cite{pu78comput,pu81expan,zhou86diff} and analytically obtains Taylor series solutions of differential equations. Although the concept of the technique is based on Taylor series expansion, it results in solving recursive algebraic equations instead of symbolically evaluating derivatives. The main focus of attention, in this note, is placed on the first order nonlinear ordinary differential equations:
\begin{equation}\label{eq.systeminitial}
  \begin{cases}
     \displaystyle \frac{\mathrm{d} x \left( t \right)}{\mathrm{d} t}= f\left( x \left( t \right),t \right) , \\
     x\left(t_0 \right) = \alpha \,.
\end{cases}
\end{equation}
The DTM leads
to representing the solution to Eq. \eqref{eq.systeminitial} in the form of a power series:
\begin{equation}\label{eq.solutiontosytem}
  x\left( t \right)=\sum\limits_{k\ge 0}{X_k{{\left( t-{{t}_{0}} \right)}^{k}}} \, ,
\end{equation}
where the unknown coefficients $X_k$ are straightforwardly evaluated by the recurrence equation:
\begin{equation}\label{eq.recurrencesytem}
  \left(k+1\right) X_{k+1}=F\left( X_k,k \right) \, , \quad k = 0,1,\cdots \,.
\end{equation}
The first coefficient, $X_0$, is assessed to be equal to the initial state $x\left(t_0 \right)$, i.e.\ $X_0 = \alpha $ (see \cite{arik08sol}) and $F\left( X_k,k \right)$ is evaluated by using the rules and techniques briefly mentioned below:
\begin{enumerate}
  \item If $f\left(t\right)=\dot{x}\left(t\right)$, then the differential transform of $f\left(t\right)$ is \[F_k=\left(k+1\right)X_{k+1}.\]
  \item If $f\left( t \right)=c\,x\left( t \right)$, then $F_k=c\,X_k$, where $c$ is a real constant.
  \item If $f\left( t \right)=x\left( t \right)\pm y\left( t \right)$, then $F_k=X_k\pm Y_k$.
  \item If $f\left( t \right)= x\left( t \right)y\left( t \right)$, then ${{F}_{k}}=\sum\nolimits_{i=0}^{k}{{{X}_{i}}{{Y}_{k-i}}}$.
  \item If $f\left( t \right)=\displaystyle \frac{x\left( t \right)}{y\left( t \right)}$, then \[{{F}_{k}}=\displaystyle \frac{1}{{{Y}_{0}}}\left( {{X}_{k}}-\sum\nolimits_{i=0}^{k-1}{{{F}_{i}}{{Y}_{k-i}}} \right), \, k \ge 1, \, F_0 = \displaystyle \frac{X_0}{Y_0}.\]
  \item $f\left( t \right)={{\left[ x\left( t \right) \right]}^{a}}$, then \[{{F}_{k}}=\sum\nolimits_{i=1}^{k}{\left( \displaystyle \frac{a+1}{k}i-1 \right) \displaystyle \frac{{{X}_{i}}}{{{X}_{0}}}{{F}_{k-i}}}, \,k \ge 1, \, F_0 = X_0^a, \, a \in \mathbb{R}.\]
  \item If $f\left( t \right)={{t}^{n}}$, then ${{F}_{k}}={{\delta }_{k-n}}$, where ${\delta }_{k-n} = \left\{ \begin{array}
      {r@{\quad \text{if} \quad}l}
      1 & k=n \\
      0 & k \ne n \,. \\
      \end{array} \right.$
  \item If $f\left( t \right)=\exp \left( \lambda t \right)$, then $F\left( k \right)=\frac{{{\lambda }^{k}}}{k!}$, $\lambda \in \mathbb{R}$.
\end{enumerate}
Proofs and more detailed descriptions can be observed in \cite{arik05sol,bervillier12status}. By evaluating the coefficients $X_k$ up to the $n$th--order, the approximate solution to Eq. \eqref{eq.systeminitial} is $x\left( t \right)=\sum\nolimits_{k=0}^{n}{{{X}_{k}}{{\left( t-{{t}_{0}} \right)}^{k}}}$.

\subsection{Adomian  decomposition method}\label{sec.DTMADM.ADM}

One of the advantages of the decomposition method is presenting analytical approximated solution to rather broad range of nonlinearities without necessitating massive numerical procedures and also restrictive assumptions in order to make the problem solvable. The method is widely used to solve problems involving algebraic, differential, integro differential, delay and partial differential equations and systems \cite{adom83Stochastic,adom86Nonlinear,adomian88review}. In order to solve Eq. \eqref{eq.systeminitial} by using ADM, the differential equation is first rewritten as follows:
\begin{equation}\label{eq.adomianform}
  Lx=g\left( t \right)+Rx+Nx \,,
\end{equation}
where $L$ denotes the first order differential operator, $R$ and $N$ represent respectively the linear and nonlinear part of $f$, and $g\left(t\right)$ denotes the remainder part of $f$ as an explicit function of $t$. Applying the inverse operator $L^{-1}$ to Eq. \eqref{eq.adomianform}, another expression of \eqref{eq.systeminitial} is obtained:
\begin{equation}
  {{L}^{-1}}\left[ Lx \right]={{L}^{-1}}\left[ g\left( t \right) \right]+{{L}^{-1}}\left[ Rx \right]+{{L}^{-1}}\left[ Nx \right]\,,
\end{equation}
where $L^{-1}$ expresses the definite integral from $t_0$ to $t$, thus:
\begin{align}\label{eq.adomianform2}
  x(t) & = \underbrace{x({{t}_{0}})+\int_{{{t}_{0}}}^{t}\!\!{g\left( t \right)\mathrm{d}t}}_{x_0} + {{L}^{-1}}\left[ Rx \right]+{{L}^{-1}}\left[ Nx \right] \nonumber \\
  & = x_0 + {{L}^{-1}}\left[ Rx \right]+{{L}^{-1}}\left[ Nx \right] \,.
\end{align}
The function $x$, which is the approximate solution to \eqref{eq.systeminitial}, and the nonlinear term $Nx$ are respectively decomposed to
\begin{equation}\label{eq.x}
  x=\sum\limits_{k\ge 0}{{{x}_{k}}}
\end{equation}
and
\begin{equation}
  Nx=\sum\limits_{k\ge 0}{{{A}_{k}}},
\end{equation}
thus Eq. \eqref{eq.adomianform2} is written in the form below:
\begin{equation}
  \sum\limits_{k\ge 0}{{{x}_{k}}} = x_0  +  {{L}^{-1}}\left[ R \sum\limits_{k\ge 0}{{{x}_{k}}} \right]  +  {{L}^{-1}}\left[ \sum\limits_{k\ge 0}{{{A}_{k}}} \right]\,,
\end{equation}
and consequently
\begin{align}
  {{x}_{1}} & = {{L}^{-1}}\left[ R{{x}_{0}} \right]+{{L}^{-1}}{{A}_{0}} \nonumber \\
  {{x}_{2}} & = {{L}^{-1}}\left[ R{{x}_{1}} \right]+{{L}^{-1}}{{A}_{1}} \nonumber \\
  \vdots \nonumber \\
  {{x}_{k+1}} & = {{L}^{-1}}\left[ R{{x}_{k}} \right]+{{L}^{-1}}{{A}_{k}}\,.
\end{align}
The polynomials $A_n$ are generated for each nonlinearity by using the formula
\begin{equation}\label{eq.adomianpoly}
  {{A}_{k}}=\frac{1}{k!}\frac{{{\mathrm{d}}^{k}}}{\mathrm{d}{{\lambda }^{k}}}{{\left[ N\left( \sum\nolimits_{i\ge 0}{{{x}_{i}}{{\lambda }^{i}}} \right) \right]}_{\lambda =0}} \,.
\end{equation}
For instance, a few terms of Adomian polynomials are as follows:
\begin{align*}
  {{A}_{0}} & = N\left( {{x}_{0}} \right) \nonumber \\
  {{A}_{1}} & = {{x}_{1}}N\left( {{x}_{0}} \right) \nonumber \\
  {{A}_{2}} & = {{x}_{2}}N\left( {{x}_{0}} \right)+\frac{1}{2!}x_{1}^{2}N\left( {{x}_{0}} \right)\nonumber \\
  {{A}_{3}} & = {{x}_{3}}N\left( {{x}_{0}} \right)+{{x}_{1}}{{x}_{2}}N\left( {{x}_{0}} \right)+\frac{1}{3!}x_{1}^{3}N\left( {{x}_{0}} \right) \,. \nonumber
\end{align*}
If the series in Eq. \eqref{eq.x} converges, the function ${{\phi }_{n}}=\sum\nolimits_{i=0}^{n}{{{x}_{i}}}$ will be the approximate analytical solution to Eq. \eqref{eq.systeminitial}.

\section{Application to the KP model}\label{sec.application}

In this section, the approximate analytical solution to KP model is obtained by application of differential transform and Adomian decomposition.
\subsection{Differential transform method}\label{sec.application.DTM}

The coefficients in the system \eqref{eq.kpScaled} are evaluated by referring to Table~3, where the initial conditions are:
\begin{equation}\label{eq.transini}
  x\left(0\right)=1 \,, \quad y\left(0\right)=1 \,, \quad z\left(0\right)=1 \,.
\end{equation}
By applying the DTM, the differential transform of the system could be obtained as follows:
\begin{equation}\label{eq.kptransform}
  \left\{ \begin{array}
  {r@{\quad=\quad}l}
   {{X}_{k+1}} & \displaystyle \frac{1}{k+1} \left( c{{Y}_{k}}-{{\mu }_{2}}{{X}_{k}}+{{p}_{1}}{{{\bar{X}}}_{k}}+{{s}_{1}}{{\delta }_{k}} \right) \,, \\
   {{Y}_{k+1}} & \displaystyle \frac{1}{k+1} \left( {{r}_{2}}{{Y}_{k}}-{{r}_{2}}b\sum\nolimits_{i=0}^{k}{{{Y}_{i}}{{Y}_{k-i}}}-a{{{\bar{Y}}}_{k}} \right) \,, \\
   {{Z}_{k+1}} & \displaystyle \frac{1}{k+1} \left( {{p}_{2}}{{{\bar{Z}}}_{k}}-{{\mu }_{3}}{{Z}_{k}}+{{s}_{2}}{{\delta }_{k}} \right) \,,
\end{array} \right.
\end{equation}
where ${\bar{X}}_k$, ${\bar{Y}}_k$ and ${\bar{Z}}_k$ are respectively equal to
\begin{equation}\label{eq.Xbar}
  {\bar{X}}_k = \left\{ \begin{array}
      {l@{\quad \text{if} \quad}l}
      \displaystyle \frac{{{X}_{0}}{{Z}_{0}}}{{{g}_{1}}+{{Z}_{0}}} & k=0 \,, \\
      \displaystyle \frac{1}{{{g}_{1}}+{{Z}_{0}}}\left( \sum\nolimits_{i=0}^{k}{{{X}_{i}}{{Z}_{k-i}}}-\sum\nolimits_{i=0}^{k-1}{{{{\bar{X}}}_{i}}{{Z}_{k-i}}} \right) & k \ge 1 \,, \\
      \end{array} \right.
\end{equation}
\begin{equation}\label{eq.Ybar}
  {\bar{Y}}_k = \left\{ \begin{array}
      {l@{\quad \text{if} \quad}l}
      \displaystyle \frac{{{X}_{0}}{{Y}_{0}}}{{{g}_{2}}+{{Y}_{0}}} & k=0 \,, \\
      \displaystyle \frac{1}{{{g}_{2}}+{{Y}_{0}}}\left( \sum\nolimits_{i=0}^{k}{{{X}_{i}}{{Y}_{k-i}}}-\sum\nolimits_{i=0}^{k-1}{{{{\bar{Y}}}_{i}}{{Y}_{k-i}}} \right) & k \ge 1 \,, \\
      \end{array} \right.
\end{equation}
\begin{equation}\label{eq.Zbar}
  {\bar{Z}}_k = \left\{ \begin{array}
      {l@{\quad \text{if} \quad}l}
      \displaystyle \frac{{{X}_{0}}{{Y}_{0}}}{{{g}_{3}}+{{Y}_{0}}} & k=0 \,, \\
      \displaystyle \frac{1}{{{g}_{3}}+{{Y}_{0}}}\left( \sum\nolimits_{i=0}^{k}{{{X}_{i}}{{Y}_{k-i}}}-\sum\nolimits_{i=0}^{k-1}{{{{\bar{Z}}}_{i}}{{Y}_{k-i}}} \right) & k \ge 1 \,, \\
      \end{array} \right.
\end{equation}
\begin{equation}
  {\delta}_k = \left\{ \begin{array}
      {r@{\quad \text{if} \quad}l}
      1 & k=0 \,, \\
      0 & k \ge 1 \, \\
      \end{array} \right.
\end{equation}
and, as stated in Sec. \eqref{sec.DTMADM.DTM}, the coefficients $X_0$, $Y_0$ and $Z_0$ are:
\begin{align*}
  X_0 & = x\left(0\right)=1 \,, \\
  Y_0 & = y\left(0\right)=1 \,, \\
  Z_0 & = z\left(0\right)=1 \,.
\end{align*}

Case 1-- No treatment ($s_1=0, s_2=0$): Without considering medical treatment, i.e.\ the external sources of effector cells and interleukin--2, $s_1$ and $s_2$, are both equal to zero, the approximate analytical solution to the system \eqref{eq.kpScaled} with initial conditions \eqref{eq.transini}, up to $O\left(t^5\right)$ is:
\begin{align}\label{eq.solDTM1}
  x\left(t\right) & = 1.0 + 0.0050622 \; t + 0.0013137 \; t^2 + 0.0005999 \; t^3 - 0.0014141 \; t^4 \,,\nonumber \\
  y\left(t\right) & = 1.0 + 0.0890891 \; t + 0.0041064 \; t^2 + 0.0001009 \; t^3 - 0.0000126 \; t^4 \,,\nonumber \\
  z\left(t\right) & = 1.0 - 5.4545450 \; t + 27.302640 \; t^2 - 91.007170 \; t^3 + 227.51860 \; t^4 \,.
\end{align}
where the immunogenicity of the tumour cells is $c=0.035$ (see Table~3). The polynomials in \eqref{eq.solDTM1} represent the solution to the KP model for a small neighbourhood of $t=0$. In order to extend the solution to a large time $T$, the DTM can be repeatedly utilized step by step, by defining a very small step--size, for instance $h = 10^{-3}$, and dividing the time $T$ into $T/h$ subintervals.

Case 2-- Immunotherapy ($s_1 >0, s_2=0$): The case of medical treatment involves the use of an external source of effectors, $s_1$, for instance \textit{lymphokine-activated killer cell} or \textit{tumor infiltrating lymphocyte}, either with or without using interleukin-2, $s_2$. For the sake of simplicity, the source of IL--2 is not considered ($s_2=0$). In the absence of IL-2, one stable non--tumour state occurs where $s_1 \ge s_{1\, crit}$. According to the values of the coefficients in Table~1, the critical value of $s_1$ is $s_{1\, crit} = 540$ (see \cite{kirsch98}). The analytical approximate solutions to the system \eqref{eq.kpScaled} with the initial conditions \eqref{eq.transini}, for $c=0.045$, $s_1=550$ and $s_2=0$, by using the DTM up to $O \left(t^5\right)$ is:
\begin{align}\label{eq.solDTM2}
  x\left(t\right) & = 1.0 + 0.0700622 \; t + 0.0007861 \; t^2 + 0.0005702 \; t^3 - 0.0013962 \; t^4 \,,\nonumber \\
  y\left(t\right) & = 1.0 + 0.0890891 \; t + 0.0011518 \; t^2 - 0.0001385 \; t^3 - 0.0000181 \; t^4 \,,\nonumber \\
  z\left(t\right) & = 1.0 - 5.4545450 \; t + 27.450370 \; t^2 - 91.500010 \; t^3 + 228.75070 \; t^4 \,.
\end{align}
\subsection{Adomian decomposition method}\label{sec.application.ADM}

For the application of the ADM to the KP system, the Adomian polynomials must be evaluated for each nonlinearity in the system. These nonlinearities are
\begin{align*}
  M_{xz} & = \frac{xz}{{{g}_{1}}+z} \,, \\
  N_{y} & = y^2 \,, \\
  P_{xy} & = \frac{xy}{{{g}_{2}}+y} \,, \\
  Q_{xy} & = \frac{xy}{{{g}_{3}}+y} \,.
\end{align*}
By using Eq. \eqref{eq.adomianpoly}, the Adomian polynomials for $M_{xz}$ and $N_y$ are calculated as below and the other two nonlinearities can be exactly evaluated similar to the nonlinearity $M_{xz}$:
\begin{eqnarray}\label{eq.nonlinearM}
  {{M}_{0}} &=& \frac{{{x}_{0}}{{z}_{0}}}{{{g}_{1}}+{{z}_{0}}} \,,  \nonumber \\
  {{M}_{1}} &=& \frac{{{x}_{0}}{{z}_{1}}+{{x}_{1}}{{z}_{0}}}{{{g}_{1}}+{{z}_{0}}} - \frac{{{x}_{0}}{{z}_{0}}{{z}_{1}}}{{{\left( {{g}_{1}}+{{z}_{0}} \right)}^{2}}}  \,, \nonumber \\
  {{M}_{2}} &=& \frac{{{x}_{0}}{{z}_{2}}+{{x}_{1}}{{z}_{1}}+{{x}_{2}}{{z}_{0}}}{{{g}_{1}}+{{z}_{0}}}  -   \frac{{{x}_{0}}z_{1}^{2}+{{x}_{1}}{{z}_{0}}{{z}_{1}}+{{x}_{1}}{{z}_{0}}{{z}_{2}}}{{{\left( {{g}_{1}}+{{z}_{0}} \right)}^{2}}}  +  \frac{z_{1}^{2}}{{{\left( {{g}_{1}}+{{z}_{0}} \right)}^{3}}} \,, \nonumber \\
  {{M}_{3}} &=& -\frac{{{x}_{0}}{{z}_{3}}+{{x}_{1}}{{z}_{2}}+{{x}_{2}}{{z}_{1}}+{{x}_{3}}{{z}_{0}}}{{{g}_{1}}+{{z}_{0}}}\nonumber \\
  & &- \frac{2{{x}_{0}}{{z}_{1}}{{z}_{2}} + {{x}_{1}}{{z}_{0}}{{z}_{2}}+{{x}_{1}}z_{1}^{2} + {{x}_{2}}{{z}_{0}}{{z}_{1}}-{{x}_{0}}{{z}_{0}}{{z}_{3}}}{{{\left( {{g}_{1}}+{{z}_{0}} \right)}^{2}}} \nonumber \\
  & &+ \frac{{{x}_{0}}z_{1}^{3}+{{x}_{1}}{{z}_{0}}z_{1}^{2} - 2{{x}_{0}}{{z}_{0}}{{z}_{1}}{{z}_{2}}}{{{\left( {{g}_{1}}+{{z}_{0}} \right)}^{3}}} +\frac{{{x}_{0}}{{z}_{0}}z_{1}^{3}}{{{\left( {{g}_{1}}+{{z}_{0}} \right)}^{4}}} \,,
\end{eqnarray}
and
\begin{eqnarray}\label{eq.nonlinearN}
  {{N}_{0}} &=& y_{0}^{2} \, \nonumber \\
  {{N}_{1}} &=& 2{{y}_{0}}{{y}_{1}} \, \nonumber \\
  {{N}_{2}} &=& y_{1}^{2}+2{{y}_{0}}{{y}_{2}} \, \nonumber \\
  {{N}_{3}} &=& 2\left( {{y}_{1}}{{y}_{2}}+{{y}_{0}}{{y}_{3}} \right)\,.
\end{eqnarray}

\begin{figure}[ht!]
\centering
\includegraphics[scale=.5]{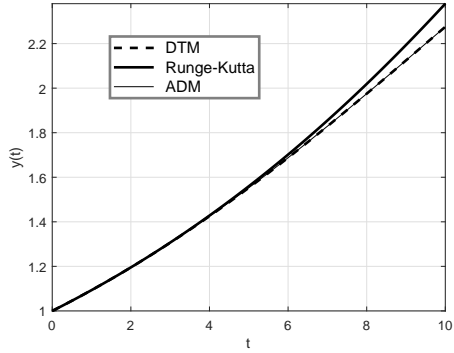}
\caption{Approximate analytical solutions to KP model by using DTM and ADM, in comparison with Runge--Kutta method, for $c=0.035$, $s_1=0$ and $s_2=0$.}
\label{fig1}
\end{figure}

Case 1-- No treatment ($s_1=0, s_2=0$): According to Eq. \eqref{eq.adomianform2}, $x_0$, $y_0$ and $z_0$ are equal to the initial state of the system \eqref{eq.kpScaled}, i.e.\
\begin{eqnarray*}
  x_0 &=& 1 \,, \\
  y_0 &=& 1 \,, \\
  z_0 &=& 1 \,,
\end{eqnarray*}
and therefore the approximate analytical solution to the KP system up to $O\left(t^5\right)$ is:
\begin{align}\label{eq.solADM1}
  x\left(t\right) & = 1.0 + 0.0050622 \; t + 0.0013136 \; t^2 + 0.0005998 \; t^3 - 0.0014141 \; t^4 \,,\nonumber \\
  y\left(t\right) & = 1.0 + 0.0890891 \; t + 0.0041063 \; t^2 + 0.0001008 \; t^3 - 0.0000126 \; t^4 \,,\nonumber \\
  z\left(t\right) & = 1.0 - 5.4545450 \; t + 27.302640 \; t^2 - 91.007170 \; t^3 + 227.51860 \; t^4 \,.
\end{align}
The solutions to \eqref{eq.kpScaled} for $y\left(t\right)$ are illustrated in Fig.\ \eqref{fig1} . The approximated analytical solutions are compared with the numerical solution to the system evaluated by using the explicit fourth order Runge-Kutta method.

\begin{figure}[ht!]
\centering
\includegraphics[scale=.5]{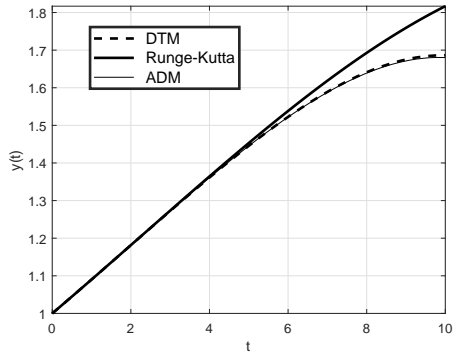}
\caption{Approximate analytical solutions to KP model by using DTM and ADM, in comparison with Runge--Kutta method, for $c=0.045$, $s_1=550$ and $s_2=0$.}
\label{fig2}
\end{figure}

Case 2-- Immunotherapy ($s_1 >0, s_2=0$): By application of the Adomian decomposition, the first terms of the series of the solution to the KP system are
\begin{eqnarray*}
  x_0 &=& 1 + \int_{0}^{t}\!\!{s_1 \mathrm{d}t} \,, \\
  y_0 &=& 1 \,, \\
  z_0 &=& 1 \,,
\end{eqnarray*}
and the solution to the system, up to $O \left(t^9\right)$ is as follows:
\begin{eqnarray}
  x\left(t\right) &=& 1.0 + 0.0700622 \; t + 0.0007861 \; t^2 + 0.0005702 \; t^3 - 0.0013962 \; t^4 \nonumber\\
   && - 5\times {{10}^{-5}} \, t^5 + 5\times {{10}^{-7}} \, t^6 + 2.9\times {{10}^{-11}} \, t^7 - 3.6\times {{10}^{-15}} \, t^8 \,, \nonumber \\
   y\left(t\right) &=& 1.0 + 0.0890891 \; t + 0.0011518 \; t^2 - 0.0001385 \; t^3 - 0.0000181 \; t^4 \nonumber\\
   && - 1\times {{10}^{-7}} \, t^5 + 4.3\times {{10}^{-9}} \, t^6 + 4.2\times {{10}^{-11}} \, t^7 - 3.2\times {{10}^{-13}} \, t^8 \,, \nonumber\\
  z\left(t\right) &=& 1.0 - 5.4545455 \; t + 27.450366 \; t^2 - 91.500006 \; t^3 + 228.75065 \; t^4 \nonumber\\
   && - 2.0877809 \, t^5 - 0.0000492 \, t^6 + 1.6\times {{10}^{-8}} \, t^7 - 4.8\times {{10}^{-12}} \, t^8 \,.
\end{eqnarray}
Figure \eqref{fig2} illustrates the approximate analytical solutions to the system \eqref{eq.kpScaled} for $y\left(t\right)$, by using the DTM and ADM compared with the numerical method of explicit Runge--Kutta.
\section{Conclusion}\label{sec.Conclusion}

Differential transform and Adomian decomposition are reliable methods of solving differential equations with fast convergence, which lead to approximate analytical solutions to a wide range of differential and integro--differential equations. In this note, a cancer immunotherapy model with rich and complicated nonlinearities has been solved by using these two methods and the results have been graphically compared with the numerical solution of the system utilizing the fourth order Runge--Kutta method.
\section*{References}

\bibliographystyle{elsarticle-num}
\bibliography{references}

\bigskip
\it
\noindent
$^1$ Department of Structure and Material\\
Universiti Teknologi Malaysia\\
81310 Skudai, Johor, Malaysia\\[4pt]
e-mail: alireza.momenzadeh@gmail.com
\hfill\\[12pt]
$^2$ Dipartimento di Scienze di Base e Applicate per l'Ingegneria \\
Sapienza Universit\`{a} di Roma \\
Via Antonio Scarpa n. 16 \\
00161 Rome, Italy \\[4pt]
e-mail: sima.sarvahrabi@sbai.uniroma1.it
\end{document}